\documentclass[twocolumn,prl,10pt,amsmath,amssymb,nofootinbib,showpacs,superscriptaddress,floatfix]{revtex4-1}

\DeclareFontFamily{U}{rcjhbltx}{}
\DeclareFontShape{U}{rcjhbltx}{m}{n}{<->rcjhbltx}{}
\DeclareSymbolFont{hebrewletters}{U}{rcjhbltx}{m}{n}

\usepackage{graphicx}
\usepackage{color}
\usepackage[usenames,dvipsnames]{xcolor}
\usepackage[colorlinks=true,linkcolor=Red,citecolor=Green,linktoc=page]{hyperref}
\usepackage{multirow}
\usepackage{float}
\usepackage{flushend}
\usepackage{balance}
\usepackage[varg]{txfonts}
\usepackage{ulem}
\usepackage{fancyhdr}

\DeclareMathSymbol{\lamed}{\mathord}{hebrewletters}{108}

\begin{document}
\title{The superconductor-insulator transition in the absence of disorder}
\author{	M.\,C.\,Diamantini}
\affiliation{NiPS Laboratory, INFN and Dipartimento di Fisica e Geologia, University of Perugia, via A. Pascoli, I-06100 Perugia, Italy}
\author{C.\,A.\,Trugenberger}
\affiliation{SwissScientific Technologies SA, rue du Rhone 59, CH-1204 Geneva, Switzerland}
\author{V.\,M.\,Vinokur}
\affiliation{Terra Quantum AG, St. Gallerstrasse 16A, CH-9400 Rorschach, Switzerland}

\begin{abstract}
We provide microscopic-level confirmation of earlier results showing that in the critical vicinity of the superconductor-to-insulator transition (SIT) disorder and localization become negligible and the structure of the emergent phases is determined by topological effects arising from the competition between two quantum orders, superconductivity and superinsulation. We find that around the critical point the ground state is a composite incompressible quantum fluid of Cooper pairs and vortices coexisting with an intertwined Wigner crystal comprising the excess of both types of excitation with respect to integer filling.
\end{abstract}

\maketitle


The superconductor-to-insulator transition (SIT)\,\cite{efetov,haviland, hebard,fisher1,fisher2, fisher3,Zant2001,girvin} 
is a paramount example of a quantum phase transition, see\,\cite{sachdev} for a review, observed in a wide variety of 2D materials, ranging from superconducting films to Josephson junction arrays (JJA), \cite{Fazio1991,Goldman1998,Goldman2010} and emerging as a key ingredient to understand high-$T_c$ superconductivity\,\cite{Chubukov2020, dtv2020}. 

Already the earliest studies of superconductivity in thin films\,\cite{efetov, Fin1983} revealed the fundamental role of the Coulomb interactions for the SIT. The observation that transport in strongly disordered films in the critical proximity of the SIT is described by the classical formulas for a regular JJA\,\cite{Fistul2008} offered a first striking evidence for self-organization of the material as an emergent granular structure independent of the original disorder. Initially, its appearance was related to localization of Cooper pairs by structural disorder\,\cite{fisher1}, viewed as a cause of the SIT. However, numerical simulations\,\cite{trivedi} demonstrated that, while the initial spatial inhomogeneity in the distribution of the superconducting order parameter reflects the effect of disorder\,\cite{sacepe1,granularity,sacepe2}, the spatial scale of the resulting granules is of order of the superconducting coherence length $\xi$ in accord with the earlier findings of\,\cite{Ioffe1981} and results of\,\cite{Fistul2008}. This confirmed that the emergent granularity, by then realized to be a crucial ingredient of the physics of the SIT, has a more universal and fundamental origin than just the disorder-induced effect.

A long-distance effective gauge theory of the SIT\,\cite{dst, dtv1,dtv2019} established the universal topological nature of the SIT and of the associated emergent granularity. However, the role of disorder remains the subject of some curiosity and surprise\,\cite{feigelman}. In this Communication we demonstrate that in the vicinity of the SIT, where the Coulomb interactions between charges are comparable with the interaction between vortices, all effects of disorder and Anderson localization are washed out by much stronger topological effects analogous to those forming the quantum Hall effect plateaus, see\,\cite{prange} for a review. Building on\,\cite{dst, dtv1,dtv2019}, we derive the microscopic ground state. We show that when the SIT occurs via an intermediate Bose (or strange) metal, the ground state is a composite topological incompressible quantum fluid of Cooper pairs and vortices coexisting with an intertwined Wigner crystal formed by the excesses of these two types of excitations with respect to integer filling. The Wigner crystal melts into a ``liquid" in a phase-separated region of coexistence  of Cooper pair- and vortex condensates when material parameters favour the direct first-order SIT.  The superconductor and superinsulator are two respective ground states hosting either a charge or a vortex condensate, depending on which of the two interactions becomes dominant. The infinite resistance of a superinsulator is caused, accordingly, by a plasma of magnetic monopole instantons, which are the 2D analogue of quantum phase slips\,\cite{dtv1}. 

To derive the emergent ground state let us first consider a similar physical setting, the quantum Hall (QH) effect. Structural disorder was long believed to be instrumental for the QH, as it was thought to provide the translation invariance breaking necessary for plateau formation by Landau levels broadening\,\cite{prange}. At the magic filling fractions corresponding to these plateaus, however, the ground states are the topological incompressible quantum fluids\,\cite{laughlin}, completely insensitive to disorder\,\cite{wenzee}. 
Furthermore, the view that structural disorder is necessary for forming QH plateaus has been recently disproved\,\cite{nodisorder}, as the necessary translation invariance breaking can be provided by the formation of a Wigner crystal in the vicinity of the magic filling fractions, so that the quantum Hall effect can take place in complete absence of disorder. The role of structural disorder is thus a secondary support as it can help to form plateaus but is not necessary for their existence. 

The process of transformation of the ground state implementing the phase change across the SIT manifests a similar phenomenon. The SIT point is universal\,\cite{fisher2} and the quantum sheet resistance $R_\square$  at this point is a longitudinal equivalent of the transverse quantum resistance of the integer QH effect. In a regime where the SIT takes place via the intermediate strange metal state\,\cite{dst, bm}, this universal point broadens to form a full-fledged intermediate topological state, the equivalent of a ``plateau", due to  topological interactions among charges and vortices. This topological state is often referred to as a Bose metal\,\cite{das} because it hosts symmetry-protected metallic edge states\,\cite{bm}. It forms due to the same competition of two quantum orders, the charge condensate and the vortex condensate that leads to the universal SIT point\,\cite{fisher2}. 

In the Bose metal, both charges, i.e. Cooper pairs, and vortices are in an out-of-condensate state because of strong quantum fluctuations. In the non-condensed state, an intertwined ensemble of charges carrying $2e$ charge and vortices carrying magnetic charges $\Phi_0=\pi/e$ (in natural units $c=1$, $\hbar =1$, $\varepsilon_0= 1$), the infrared-dominant interactions are the mutual statistics Aharonov-Casher-Bohm interactions stemming from the quantum phases acquired by mutually encircling charges and vortices.  All other interactions, including localization effects, become negligible in the long distance limit. As pointed out by Wilczek\,\cite{wilczek}, a local representation of these mutual statistics interactions is achieved by introducing two fictitious gauge fields $a_{\mu}$ and $b_{\mu}$ with the mixed Chern-Simons interaction\,\cite{jackiw}, 
\begin{equation}
	{\cal L} = {1\over 2\pi} a_{\mu} \epsilon^{\mu \alpha \nu} \partial_{\alpha} b_{\nu} - a_{\mu} j^{\mu}- b_{\mu} \phi^{\mu} \ ,
	\label{mixed}
\end{equation}
where $j^{\mu}$ and $\phi^{\mu}$ are charge and vortex currents, respectively. 

To proceed towards a microscopic picture we adopt the technique of\,\cite{higgsless} and consider vortices of the two opposite signs as living on two distinct fictitious planes that we will solder together at the end. Let us begin by considering `up' vortices only. 
In this case, the time-reversal symmetry is broken. It will be restored later when we combine the two chiralities.  
We adopt the usual mean field approximation of anyon physics,  see\,\cite{wilczekbook} for a review, by replacing the homogeneously distributed vortices by the uniform magnetic field ${\cal B}_{\Phi} = 2e N_{\Phi} \Phi_0 /A $ felt by the Cooper pairs, where $N_{\Phi}$ is the number of vortices and $A$ the area of the sample, and the charge distribution with the corresponding uniform ``magnetic field" 
${\cal B}_{Q} = \Phi_0 N_Q 2e /A $ felt by the vortices via the Magnus force, the dual of the Lorentz force, see, e.g.\,\cite{magnus}. The hydrodynamic Magnus force is experienced by a vortex moving with the velocity  ${\bf v}$ over the charge background, and in two dimensions (2D) becomes
\begin{equation}
	F_{\rm M}^i = \kappa \epsilon^{ij} v^j \rho
	\label{magnus1}
\end{equation} 
where $\rho$ is the fluid density and $\kappa$ is the vorticity, the circulation of the vortex quantized in multiples of $2\pi$. 
When the fluid is charged (an elemental charge is $2e$)  it can be written as 
\begin{equation}
	F_{\rm M}^i = \Phi  \epsilon^{ij} v^j 2e \rho\ ,
	\label{magnus2}
\end{equation} 
where $\Phi$ is the vorticity normalized in units of $\Phi_0$. This is the exact dual of the usual Lorentz force $F_{\rm L}^i =
2e \epsilon^{ij} v_j B$, from which we evince that the charge density plays for vortices exactly the same role as the magnetic field for charges. 

At the self-dual point, where $N_{\mathrm Q} = N_\Phi$, the two ``magnetic fields" ${\cal B}_{\mathrm Q}$ and ${\cal B}_\Phi$ become identical,
${\cal B}_{\mathrm Q} ={\cal B}_\Phi = {\cal B}$. At this point we use the known results for the quantum state of two flavours of particles with Coulomb interactions and mutual statistics in a magnetic field\,\cite{macdonald, wilczek} to write down the wave-function of the incompressible quantum fluid of charges and vortices at the self-dual point. In the general case it is
\begin{eqnarray}
	\Psi_{m_1, m_2, n} \left( \{ z_i\} , \{ w_i \} \right) = \prod_{i,j} (z_i-z_j )^{m_1} \prod_{i,j} (w_i-w_j )^{m_2} \prod_{i,j} (z_i-w_j )^{n} 
	\nonumber \\
	\times \ {\rm e}^{ -{{\cal B} \over 4} \left( \sum_i |z_i|^2 + |w_i|^2 \right) } \ ,
	\label{general}
\end{eqnarray}
where $z_i$ denote the coordinates of the Cooper pairs and $w_i$ those of vortices and $m_1$ and $m_2$ are even integers to guarantee the bosonic statistics of Cooper pairs and vortices. The total filling fraction (summing over both flavours) $\nu$ and the torus degeneracy $D_{\mathrm T}$ are given by\,\cite{wilczek}, 
\begin{eqnarray}
	\nu = {m_1 + m_2 -2n\over m_1 m_2 -n^2} \ ,
	\nonumber \\
	D_{\mathrm T} = | m_1 m_2 -n^2 | \ .
	\label{par}
\end{eqnarray}
Restricting to the integer case (for both flavours) requires setting $m_1 = m_2 = 0$ and $n=1$. In this case there is no topological order, encoded in a non-trivial torus degeneracy and the mutual statistics is also integer. It turns, however, out that  the wave function in the integer case cannot be obtained by simply setting $m_1=0$ and $m_2=0$ in the above formula since this state would be unstable with respect to phase separation\, \cite{senthillevin}. The correct wave function of the topological incompressible fluid of the two boson system of Cooper pairs and up vortices with $U(1)$ symmetry protected gapless edge states at the SIT self-dual point is\,\cite{senthillevin}, 
\begin{equation}
	\Psi \left( \{ z_i\} , \{ w_i \} \right) = \prod_{i<j} |z_i-z_j| \prod_{i<j}|w_i-w_j| 
	\prod_{i,j} {(z_i-w_j )\over |z_i-w_j|} \ {\rm e}^{ -{{\cal B} \over 4} \left( \sum_i |z_i|^2 + |w_i|^2 \right) } \ .
	\label{special}
\end{equation}

Exactly as in the analogous quantum Hall effect situation\,\cite{nodisorder}, small deviations from the self-dual point lead to small excess densities of quasi-particle and quasi-hole excitations of the two kinds. It comes at no surprise that, for an integer topological state, these excitations are charges and vortices themselves\,\cite{wilczek}. On the superconducting side there will be an excess of charges and an excess of vortices on the insulating side. And, as pointed out in\,\cite{nodisorder}, for sufficiently small densities near the self-dual point, these excitations behave as classical particles with fixed guiding-centers in the respective magnetic fields. Since both have long-range Coulomb interactions, they form a Wigner crystal for sufficiently low temperatures. 

The exact same wave function can be formulated for the `down' vortices on the second fictitious plane, with the only difference that it has the opposite chirality. Soldering the two fictitious planes means taking their ``singlet quantum superposition", i.e.
building a quantum superposition of conjugate `up' and `down' vortex configurations at the same locations with the relative (-1) sign\,\cite{xu}. This wave function describes the simplest bosonic topological insulator\,\cite{lu}, with the long-distance effective Lagrangian
\begin{equation}
	{\cal L} = {1\over 2\pi} c_{\mu} \epsilon^{\mu \alpha \nu} \partial_{\alpha} d_{\nu} \ ,
	\label{topins}
\end{equation}
expressed in terms of two emergent gauge fields $c_{\mu}$ and $d_{\mu}$, a vector and a pseudovector so that parity and time reversal symmetries are preserved\,\cite{dst}. Edge modes of this model mediate an electric current leading to the quantum resistance $R_\square$$=$$R_{\mathrm Q}$ at the self-dual point\,\cite{bm}. 

The ground state near the self-dual point 
is, therefore, a composite topological incompressible quantum fluid of Cooper pairs and vortices. Excess charges and vortices, with respect to  integer filling, form a Wigner crystal
of guiding centers of singlet quantum superpositions of the two possible rotation chiralities. When the charge excitation density becomes sufficiently large and achieves a certain threshold value, the charges condense via a quantum Berezinskii-Kosterlitz-Thouless transition\,\cite{ber,kos}, and a superconductor forms. The same happens with vortices on the other side of the self-dual point, with the condensation giving rise to a superinsulator\,\cite{bm}. 

In the range of parameters favoring the formation of an intermediate Bose metal plateau the observed granular structure\,\cite{granularity, sacepe1, sacepe2} is a regular intertwined Wigner crystal of charges and vortices. Since in the presence of even weak disorder, the Wigner crystal remains pinned as long as the applied electric current or voltage do not exceed a critical value\,\cite{Drichko2009}, electric conduction is mediated exclusively by the edge modes of the topological state. The only difference with respect to the QH effect case is that the creation of a small amount of quasi-particles slightly modifies the two ``magnetic fields" felt by the two flavours of excitations. This effect is balanced by effective couplings $e^{\prime}$ and $\Phi_0^{\prime}$ so that
\begin{equation}
	e^{\prime} N_{\Phi}^{\prime} \Phi_0 = e N_{\mathrm Q}^{\prime} \Phi_0^{\prime} \ ,
	\label{balance1} 
\end{equation}
which can be rewritten as
\begin{equation}
	{e^{\prime} \over e} ={ N_{\mathrm Q}^{\prime} \Phi_0^{\prime} \over N_{\Phi}^{\prime} \Phi_0} \ .
	\label{balance2}
\end{equation}
If we require that the Dirac quantization condition, i.e. the integer mutual statistics, is maintained we will thus have the same
topological quantum state but with an effective, renormalized coupling constant
\begin{equation}
	e \to e^{\prime} = e\sqrt{g}\,, \,\,\,\,\,\,\,
	g ={N_{\mathrm Q}^{\prime} \over N_{\Phi}^{\prime} } \ ,
	\label{renor}
\end{equation}
which leads to a renormalized resistance 
\begin{equation}
	R_{\square} = {R_{\rm Q} \over g} \ ,
	\label{resistance}
\end{equation}
around the self-dual point, an effect that has been experimentally confirmed\,\cite{bm}. 

When parameters favor a direct SIT, which is a first-order phase transition without an intermediate Bose metal, charge and vortex condensates coexist within a finite range around the self-dual point\,\cite{dst, bm}. In this case, we have a ``liquid" granular structure consisting of intertwined bubbles of superconducting and superinsulating condensates. This is the phase separation state of two boson flavors discussed in\,\cite{senthillevin}.
The exact parameters of the Wigner crystal will be analyzed in a forthcoming publication. Here we would like to point out that the formation of a granular structure in the SIT emerges mainly from the topological interactions arising from the competition of two quantum orders. 

\subsection{Acknowledgements}
	The work by V.M.V. was supported by Terra Quantum.
	M.C.D. thanks CERN, where she completed this work, for kind hospitality.
\bigskip

\subsection{Data availability}
Data sharing is not applicable to this article as no datasets were generated or analyzed during the current study.




\end{document}